\documentclass{elsart}
 \usepackage{graphicx}
 \usepackage{bm}
 \usepackage{amssymb}
 \usepackage[mathcal]{euscript}

\journal{Journal of Computational Physics}

\begin{document}

\begin{frontmatter}

\title{Going forth and back in time\,:\\ a fast and parsimonious algorithm for
mixed initial/final-value problems}

\author[inln,oca]{Antonio Celani},
\author[oca,uroma]{Massimo Cencini},
\author[oca]{Alain Noullez\corauthref{cra}}
\corauth[cra]{Corresponding author.}
\ead{anz@obs-nice.fr}

\address[inln]{CNRS, INLN, 1361 Route des Lucioles, F-06560 Valbonne, France}
\address[oca]{CNRS, Observatoire de la C\^ote d'Azur,
B.P. 4229, F-06304 Nice Cedex 4, France}
\address[uroma]{Dipartimento di Fisica Universit\`a di Roma ``La Sapienza'' and
Center for Statistical Mechanics and Complexity INFM UdR Roma 1,
Piazzale Aldo Moro, 2, I-00185 Roma, Italy}

\begin{abstract}

We present an efficient and parsimonious algorithm to solve mixed
initial/final-value problems.  The algorithm optimally limits the memory
storage and the computational time requirements\,: with respect to a simple
forward integration, the cost factor is only logarithmic in the number of
time-steps.  As an example, we discuss the solution of the final-value problem
for a Fokker-Planck equation whose drift velocity solves a different
initial-value problem -- a relevant issue in the context of turbulent scalar
transport.

\end{abstract}

\begin{keyword}
 Initial/final-value problems \sep turbulent transport
 \MSC 65M99 \sep 76F25
\end{keyword}

\end{frontmatter}


\section{Introduction}
\label{sec1}

In the investigation of dynamical systems, the standard initial-value problem
is to compute from the equations of motion the state of the system at a final
time~$t$, given its initial condition at time~$t_0$.  Sometimes, however, the
state of the system might be known at a final time~$t$, and one would be
interested in evolving the system backward in time to compute its earlier
states, back to~$t_0$.  This can in theory be easily accomplished by reversing
the direction of the time-integration, thus transforming the final-value
problem in an initial-value one.  Problems might however appear if the forward
evolution is given by a mapping that is not one-to-one, as the previous state
can thus become undefined.  Even if the time evolution is given by a
differential system, the backward evolution becomes unstable if the forward
dynamics is dissipative, as is the case in many physical systems, like for
instance Navier-Stokes turbulence.  The problem stems from the fact that a
dissipative system contracts volumes in phase space in the forward direction,
and thus expands them in the backward direction and amplifies any small
numerical errors, like those caused by roundoff.

Another quite difficult task is to obtain the evolution of a system when part
of the variables that specify the state are given at an initial time~$t_0$ and
the remaining ones are given at the final time~$t$.  We refer to this class of
problems as ``mixed initial/final-value''.  We will be interested in a special
subclass of such problems, that can be schematically written as follows
\begin{eqnarray}
 {\d {\bm u} \over \d s} &=& {\bm f}({\bm u},s)\,, \qquad\,\;\;\;\;
 {\bm  u}(t_0) \!= {\bm u}_0\;\label{eq:0.1}\\
 {\d {\bm z} \over \d s} &=& {\bm g}({\bm z},{\bm u},s)\,, \qquad\;
 {\bm z}(t) = {\bm z}_t\;,\label{eq:0.2}
\end{eqnarray}
where~${\bm u}$ and~${\bm z}$ are vectors in a given space.  Far from being an
academic oddity, this problem is relevant to many physical situations, among
which we will discuss in detail the transport of scalar fields by a dynamically
evolving flow.  Consider indeed the problem of finding the solution~${\bm
a}(s)$ of the stochastic differential equation
\begin{equation}
 {\d{\bm a}(s) \over \d s} = {\bm v}({\bm a}(s),s)
                             +\sqrt{2\kappa}\,{\bm \eta}(s)\;,
 \label{eq:1}
\end{equation}
with the final value~${\bm a}(t) = {\bm x}$.

Eq.~(\ref{eq:1}) describes the evolution of a particle transported by the
velocity field~${\bm v}$ and subject to molecular diffusion with
diffusivity~$\kappa$, represented here by the zero-mean Gaussian process~${\bm
\eta}$ with correlations~$\langle \eta_i(t)\eta_j(t') \rangle =
\delta_{ij}\delta(t-t')$.  The velocity field~${\bm v}$ at any time~$s$ has to
be obtained from some dynamical law (e.g. the Navier-Stokes equations) and from
its initial value at~$s = t_0$. It is easy to recognize that ${\bm v}$~plays
the role of the variable~${\bm u}$ in Eqs.~(\ref{eq:0.1}-\ref{eq:0.2}) whereas
${\bm a}$~has to be identified with~${\bm z}$.  An equivalent description may
be given in terms of the transition probability $P({\bm y},s|{\bm x},t)$ --
i.e. the probability that a particle is in~${\bm y}$ at time~$s$ given that it
will be in~${\bm x}$ at time~$t$.  The propagator evolution is ruled by the
well known Kolmogorov equation~\cite{R96,G96}
\begin{equation}
 -\partial_s P({\bm y},s|{\bm x},t) - {\bm \nabla}_{\!y} \cdot
 [{\bm v}({\bm y},s) P({\bm y},s|{\bm x},t)] =
 \kappa\, \nabla^2_y P({\bm y},s|{\bm x},t) \;,
 \label{eq:4.2}
\end{equation}
where the final condition is set\,: $P({\bm y},t|{\bm x},t) = \delta({\bm
x}-{\bm y})$.  In the latter case, it is~$P$ that has to be interpreted
as~${\bm z}$ in~(\ref{eq:0.2}).

In this article, we propose a fast and memory-sparing algorithm to solve the
problem~(\ref{eq:0.1}-\ref{eq:0.2}) or to allow one to go back through the time
evolution if the dynamics is unstable or non-invertible.  In Sec.~\ref{sec:2}
we describe in detail the algorithm, comparing it to more naive and less
efficient strategies.  In Sec.~\ref{sec:3} we present an application to the
problem of front generation in passive scalar turbulence (see
e.g.~\cite{SS00,FGV01}).


\section{Backward Algorithm}
\label{sec:2}

The obvious difficulty with Eqs.~(\ref{eq:0.1}--\ref{eq:0.2}) resides in the
fact that, since the initial conditions of~${\bm u}$ and~${\bm z}$ are set at
different times, they cannot be evolved in parallel.  Also, the time evolution
of~${\bm u}(s)$ might be non-invertible or unstable in the backward time
direction.  The whole history of~${\bm u}(s)$ in the interval~$[t_0,t]$ is thus
needed to integrate~${\bm z}(s)$ from time~$t$ back to time~$t_0$.

Before presenting our own algorithm, we wish to discuss some naive strategies
to expose their shortcomings and advantages, and introduce notations.  In the
following, we will assume the whole time interval~$[t_0,t]$ to be discretized
in $N$~identical time steps, small enough to ensure accurate integration of
Eqs.~(\ref{eq:0.1}--\ref{eq:0.2}).  The states~${\bm u}(s)$, ${\bm z}(s)$ thus
have to be computed at the $N+1$~times $t_0,\ldots,t_j \equiv
t_0+(t-t_0)j/N,\ldots,t_N \equiv t$.  In our applications, the states~${\bm
u}(s)$ and~${\bm z}(s)$ will be $d$-dimensional vector fields, numerically
resolved with $\mathcal{L}^d$ collocation points, and therefore have a
size~$O(d\,\mathcal{L}^d)$, typically very large, that will be taken as unit of
measure when describing the storage requirements~$\mathcal{S}(N)$ of the
different algorithms.  The CPU~time cost~$\mathcal{T}(N)$ will refer only to
(forward) integrations of~${\bm u}$ and will be expressed in terms of the time
to perform a single forward integration step.  We will also give examples of
memory use and CPU~time for~$d = 2$, $\mathcal{L} = 1024$ and~$N = 2^{14}$,
which are typical values of moderately resolved direct numerical simulations in
computational fluid dynamics, requiring 16~MB of memory to store a single state
array.

The most obvious and simple strategy is the following\,:
\begin{description}
 \item{\bf A1.} integrate forward Eq.~(\ref{eq:0.1}) from~$t_0$ to~$t_N$ and
store~${\bm u}(s)$ at all time steps~$t_0 \ldots t_N$\,;
 \item{\bf A2.} integrate backward Eq.~(\ref{eq:0.2}) from~$t_N$ back to~$t_0$.
\end{description}
The number of integration steps needed by this procedure is~$\mathcal{T}(N) =
N$, while the memory storage cost is a frightening~$\mathcal{S}(N) = N$.  As
soon as the dimensionality of the space or the number of collocation points
increase, this approach becomes rapidly unfeasible.  Taking our typical fluid
dynamics value, one would need 256~GB of memory, which is clearly irrealistic.

A different strategy that minimizes the memory requirements is\,:
\begin{description}
 \item{\bf B1.} set~$n \leftarrow N$ and store the initial
condition~${\bm u}_0$\,;
 \item{\bf B2.} integrate forward Eq.~(\ref{eq:0.1}) from~$t_0$ to~$t_n$\,;
 \item{\bf B3.} integrate backward Eq.~(\ref{eq:0.2}) from~$t_n$ to~$t_{n-1}$,
update~$n \leftarrow n-1$, and go back to step~{\it B2} if~$n \ge 0$.
\end{description}
While this method is very advantageous in memory~$\mathcal{S}(N) = 1$, it is
prohibitively expensive because of the large number of iterations
needed\,:~$\mathcal{T}(N) = N(N+1)/2$.  With the previously given numerical
parameters, one needs a daunting increase by a factor~$8200$ in CPU~time with
respect to algorithm~{\it A}.

To improve algorithm~{\it B}, one can think of using more memory and a simple
generalization goes as follows\,:
\begin{description}
 \item{\bf C1.} integrate forward Eq.~(\ref{eq:0.1}) from~$t_0$ to~$t_N$ and
store the states~${\bm u}(s)$ at the $M$~equidistant times $\tau_k =
t_{Nk/M},\; k = 0,\ldots,M-1$ (we assume here $N$~to be a multiple of~$M$ for
convenience)\,;
 \item{\bf C2.} apply algorithm~{\it B} successively in each
segment~$[\tau_k,\tau_{k+1}]$.
\end{description}
The number of operations is~$\mathcal{T}(N) = N(N+M)/(2M)$ remains however
prohibitive unless we raise~$M$ to be~$O(N)$.  Now, since the memory storage
is~$\mathcal{S}(N) = M$, $M$~cannot be made too large as well.  Again refering
to the numerical parameters given above, we have that for~$M = 16$ the storage
requirement is reasonably low (256~MB) yet the time factor with respect to
algorithm~{\it A} is a still discouraging~512.

A further possibility which helps reducing the number of iterations and is
almost reasonable for the memory storage needs is\,:
\begin{description}
 \item{\bf D1.} integrate forward Eq.~(\ref{eq:0.1}) from~$t_0$ to~$t_N$ and
store $M$~states~${\bm u}(s)$ at times~$\tau_k = t_{Nk/M},\; k =
0,\ldots,M-1$. Set~$k \leftarrow M-1$\,;
 \item{\bf D2.} integrate forward Eq.~(\ref{eq:0.1}) from~$t_{Nk/M}$
to~$t_{N(k+1)/M}$, using the stored state at~$\tau_k$ as initial condition and
saving the states~${\bm u}(s)$ at all time steps in a further set of
$N/M$~storage locations\,;
 \item{\bf D3.} integrate backward Eq.~(\ref{eq:0.2}) from~$t_{N(k+1)/M}$
to~$t_{Nk/M}$ using the $N/M$~saved~${\bm u}(s)$, update~$k \leftarrow k-1$ and
go back to step~{\it D2} if~$k \ge 0$.
\end{description}
This procedure needs a reasonable total number of time steps~$\mathcal{T}(N) =
N+(N-M)(M-1)/M$, that is~$\mathcal{T}(N) \approx 2N$ when~$N \gg M \gg 1$, and
is thus asymptotically linear in~$N$, provided we have enough memory.  The
storage requirement is indeed rather large~$\mathcal{S}(N) = M+N/M$ and is
minimized for a fixed~$N$ by taking~$M = \sqrt{N}$.  With our typical
parameters, we would have to take~$M = 128$ and store 256~fields, amounting to
roughly~4~GB, still too large for typical workstations.

Algorithm~{\it D} is still too greedy in memory, but it gives the idea of
dividing the problem into smaller subproblems that have a much smaller running
time, and that be combined later to give the full solution.  If we push this
idea further, we can build a recursive algorithm that integrates backward
from~$t_N$ to~$t_0$ by integrating forward from~$t_0$ to~$t_{N/2}$, and using
the states at~$t_0$ and~$t_{N/2}$ to call itself successively in the
intervals~$[t_N,t_{N/2}]$ and~$[t_{N/2},t_0]$.  We have chosen here the
subdivision base (the equivalent of~$M$ in the previous algorithm) to be~2,
because it gives the simplest and one of the most efficient algorithms, but
other bases could be used to slightly reduce the number of integration steps,
at the price of using more storage.  Of course, recursion can be eliminated and
it is in its non-recursive form that we will describe our procedure.  To do
that, we will need a {\em stack}, that is a list of states to which we can add
(push/save) a new item or remove (pull/delete) the last stored item.  A stack
can always be implemented as an array in programming languages that do not have
it as a built-in type.  We will also use the index~$[{\rm top}]$ to refer to
the (last pushed) element on top of the stack.  Our algorithm is then very easy
to state\,:
\begin{description}
 \item{\bf R1.} set the desired time index~$n \leftarrow N$ and push the
initial condition~${\bm u}_0$ on the stack\,;
 \item{\bf R2.} if the state on top of the stack does not correspond to the
index~$n$, set~$j \leftarrow (j_{[{\rm top}]}+n+1)/2$ to the upper midpoint of
the interval, integrate forward~${\bm u}_{[{\rm top}]}$ from~$t_{[{\rm top}]}$
to~$t_j$, push the state~${\bm u}_j$ on the stack and go back to
step~{\it R2}\,;
 \item{\bf R3.} pull the state~${\bm u}_{[{\rm top}]} \equiv {\bm u}_n$ from
the stack, use it to integrate backward Eq.~(\ref{eq:0.2}) from~$t_n$
to~$t_{n-1}$, set~$n \leftarrow n-1$, and go back to step~{\it R2} if~$n \ge
0$.
\end{description}

\begin{figure}[t]
 \def\bs#1{\scriptsize\quad\begin{tabular}{|#1|}\hline}
 \def\es{\\\hline\end{tabular}{\quad}}
 \def\u#1{${\bm u}_{#1}$}
 \def\tt#1{$s = t_{#1}$}
 \begin{center}
 \begin{tabular}{|rlrr|}\hline
 \multicolumn{1}{|c|}{Time} & \multicolumn{1}{c|}{Stack} &
 \multicolumn{1}{c|}{Steps} & \multicolumn{1}{c|}{Total steps}\\ \hline
\tt{20}& \bs{c|c|c|c|c|c} \u{0}& \u{10}& \u{15}& \u{18}& \u{19}& \u{20} \es &
20 & 20 \\
\tt{19}& \bs{c|c|c|c|c} \u{0}& \u{10}& \u{15}& \u{18}& \u{19} \es & 0 & 20 \\
\tt{18}& \bs{c|c|c|c} \u{0}& \u{10}& \u{15}& \u{18} \es & 0 & 20 \\
\tt{17}& \bs{c|c|c|c|c} \u{0}& \u{10}& \u{15}& \u{16}& \u{17} \es & 2 & 22 \\
\tt{16}& \bs{c|c|c|c} \u{0}& \u{10}& \u{15}& \u{16} \es & 0 & 22 \\
\tt{15}& \bs{c|c|c} \u{0}& \u{10}& \u{15} \es & 0 & 22 \\
\tt{14}& \bs{c|c|c|c|c} \u{0}& \u{10}& \u{12}& \u{13}& \u{14} \es & 4 & 26 \\
\tt{13}& \bs{c|c|c|c} \u{0}& \u{10}& \u{12}& \u{13} \es & 0 & 26 \\
\tt{12}& \bs{c|c|c} \u{0}& \u{10}& \u{12} \es & 0 & 26 \\
\tt{11}& \bs{c|c|c} \u{0}& \u{10}& \u{11} \es & 1 & 27 \\
\tt{10}& \bs{c|c} \u{0}& \u{10} \es & 0 & 27 \\
\tt{9}& \bs{c|c|c|c|c} \u{0}& \u{5}& \u{7}& \u{8}& \u{9} \es & 9 & 36 \\
\tt{8}& \bs{c|c|c|c} \u{0}& \u{5}& \u{7}& \u{8} \es & 0 & 36 \\
\tt{7}& \bs{c|c|c} \u{0}& \u{5}& \u{7} \es & 0 & 36 \\
\tt{6}& \bs{c|c|c} \u{0}& \u{5}& \u{6} \es & 1 & 37 \\
\tt{5}& \bs{c|c} \u{0}& \u{5} \es & 0 & 37 \\
\tt{4}& \bs{c|c|c|c} \u{0}& \u{2}& \u{3}& \u{4} \es & 4 & 41 \\
\tt{3}& \bs{c|c|c} \u{0}& \u{2}& \u{3} \es & 0 & 41 \\
\tt{2}& \bs{c|c} \u{0}& \u{2} \es & 0 & 41 \\
\tt{1}& \bs{c|c} \u{0}& \u{1} \es & 1 & 42 \\
\tt{0}& \bs{c} \u{0} \es & 0 & 42 \\[4pt] \hline
 \end{tabular}
 \end{center}
 \caption{Algorithm~{\it R} in action for~$N = 20$.  The state of the stack is
shown for every time~$t_j$ at the beginning of label~{\it R3}, where the
state~${\bm u}(t_j)$ becomes available on top of the stack.  {\em Steps\/} is
the number of forward integration steps needed for this particular time, while
{\em Total steps\/} is the number of forward steps since the final
time~$t_{20}$ to the current time.}
 \label{figa}
\end{figure}

To understand better the behavior of algorithm~{\it R}, the easiest is to show
an example of how it works in a simple case for a small value of~$N$. 
Fig.~\ref{figa} does this for~$N = 20$, showing the stack movements at every
time step.  Even for such a small value of~$N$, algorithm~{\it R} needs
3.5~times less memory and is only 2.1~times slower than algorithm~{\it A},
while it is 5~times faster than algorithm~{\it B}.

It is obvious that our algorithm needs a very small amount of
storage~$\mathcal{S}(N) = 1+\lceil\log_2(N)\rceil$, that is only 15~fields
or~240~MB for our typical example with~$N = 2^{14}$.  The computing time is
also very reasonable\,: the computing time~$\mathcal{T}(N)$ obeys the
recursions~$\mathcal{T}(N) = 2\,\mathcal{T}(N/2)+N/2-1$ if~$N$ is even
and~$\mathcal{T}(N) = 2\,\mathcal{T}(\lfloor N/2 \rfloor)+\lceil N/2 \rceil$ if
it is odd.  The number of steps thus depends on the precise binary
representation of~$N$, but is given approximately by~$\mathcal{T}(N) \approx
N\lceil\log_2(N+1)\rceil/2+1$ (equality being achieved if~$N$ is a power of~2),
that is a cost factor that is only logarithmic in the number of time steps.  In
our same example, we find that we will need 7~times more integration steps than
the brute force algorithm~{\it A}, but 1100~times less storage, so that the
whole stack can be kept in-core during the backward integration.

The algorithm we propose is thus quite efficient in computing time, and very
economical in memory, opening the door to the study of the backward evolution
of very large multi-dimensional fields.  To give an idea of possible
applications, one might study the ``seed'' at~$t_0$ that gave birth to a
particular structure observed at time~$t$.  As an example, we will discuss in
the following section the numerical implementation and an application of this
algorithm to scalar transport in turbulent flows.


\section{Scalar fields in turbulent flows}
\label{sec:3}

The transport of scalar fields, such as temperature, pollutants and chemical or
biological species advected by turbulent flows, is a common phenomenon of great
importance both in theory and applications~\cite{SS00}.  A scalar field,
$\theta({\bm x},t)$, obeys the advection-diffusion equation
\begin{equation}
 \partial_t \theta + {\bm v}\cdot{\bm \nabla}\theta =
 \kappa\nabla^2\,\theta + \phi \;,
 \label{eq:5}
\end{equation}
where $\kappa$~is the molecular diffusivity, ${\bm v}$~is the velocity field,
and~$\phi$ is the scalar input acting at a characteristic lengthscale~$L_\phi$.
The presence of a scalar source allows for studying stationary properties. 
Thanks to the linearity of Eq.~(\ref{eq:5}), the problem can be solved in terms
of the particle propagator~\cite{SS00,FGV01}
\begin{equation}
 \theta({\bm x},t) = \int_0^t\!\!\d s \int\!\!\d{\bm y}\;
 P({\bm y},s|{\bm x},t)\,\phi({\bm y},s)\;,
 \label{eq:7}
\end{equation}
as can be directly checked by inserting~(\ref{eq:7}) in~(\ref{eq:5}) and
using~(\ref{eq:4.2}).  To make more intuitive the physical content of
Eq.~(\ref{eq:7}), we can rewrite it as
\begin{equation}
 \theta({\bm x},t) =
 \left\langle \int_0^t\!\!\d s\; \phi({\bm a}(s),s) \right\rangle_{\bm a}\;,
 \label{eq:7bis}
\end{equation}
where $\langle\ldots\rangle_{\bm a}$ denotes the average over particle
trajectories obeying~(\ref{eq:1}) with ${\bm a}(t) = {\bm x}$.
From~(\ref{eq:7bis}) one understands that~$\theta({\bm x},t)$ is built up by
the superposition of the input along all trajectories ending at point~${\bm x}$
at time~$t$.

The velocity field evolves according to the Navier-Stokes equation\,:
\begin{equation}
 \partial_t{\bm v} + {\bm v}\cdot{\bm \nabla}{\bm v} =
 -{\bm \nabla}p + \nu\nabla^2{\bm v} + {\bm f} \;.
 \label{eq:6}
\end{equation}
Where the pressure~$p$ is fixed by the incompressibility condition (${\bm
\nabla}\cdot{\bm v} = 0$), $\nu$ is the kinematic viscosity, and~${\bm f}$ the
energy input. Notice that $\theta$~does not enter the equation for the velocity
field and therefore the scalar is called passive.

In the following we will consider a passive scalar field evolving in a two
dimensional turbulent velocity field, and show how the numerical study of
particle propagator conveys some information on the dynamical origin of
structures in the scalar field.


\subsection{Numerical Implementation}

We integrate Eqs.~(\ref{eq:5}), (\ref{eq:6}) and~(\ref{eq:4.2}) in a doubly
periodic box $2\pi \times 2\pi$ with $\mathcal{L}_x \times \mathcal{L}_y$ grid
points (the results here discussed are for $\mathcal{L}_x = \mathcal{L}_y =
1024$) by a standard $2/3$-dealiased pseudo-spectral
method~\cite{Orszag,CHQZ88}.  A detailed description of the properties of the
velocity field in two-dimensional Navier-Stokes turbulence can be found in
\cite{BCV00}.  Here we only mention that the velocity field is self-similar
with Kolmogorov scaling i.e. $({\bm v}({\bm x}+{\bm r},t)-{\bm v}({\bm
x},t))\cdot{\bm r}/r \sim r^{1/3}$.  However, the passive scalar increments
$\theta({\bm x}+{\bm r})-\theta({\bm x})$ are not self-similar, since large
excursions occur with larger and larger probability for increasingly small
separations~$r$ (see, e.g., \cite{CLMV00,CLMV01}).

Time integration of Eqs.~(\ref{eq:5}) and~(\ref{eq:6}) is performed using a
second order Runge-Kutta scheme modified to integrate exactly the dissipative
terms.  Both the velocity field and passive scalar were initialized to zero and
integrated for a transient until a statistically stationary state was reached.
The propagator is initialized at the final time as a Gaussian $P({\bm y},t|{\bm
x},t) = \exp[-|{\bm x}-{\bm y}|^2/(2\delta^2)]/(\sqrt{2\pi}\delta)$, where the
width~$\delta$ is of the order of few grid points The time evolution of
Eq.~(\ref{eq:4.2}) is implemented by a second-order Adams-Bashforth scheme
modified to exactly integrate the dissipative terms. The adoption of different
schemes for the forward and backward integration is motivated by the
requirement of minimizing the use of Fast Fourier Transforms.  To implement the
backward algorithm it is also necessary to store the scalar and velocity
forcings, and this is easily accomplished by including in the stored states the
seed(s) of the pseudo-random number generator(s).

\begin{figure}[t]
 \centerline{
 \includegraphics[scale=.6,draft=false]{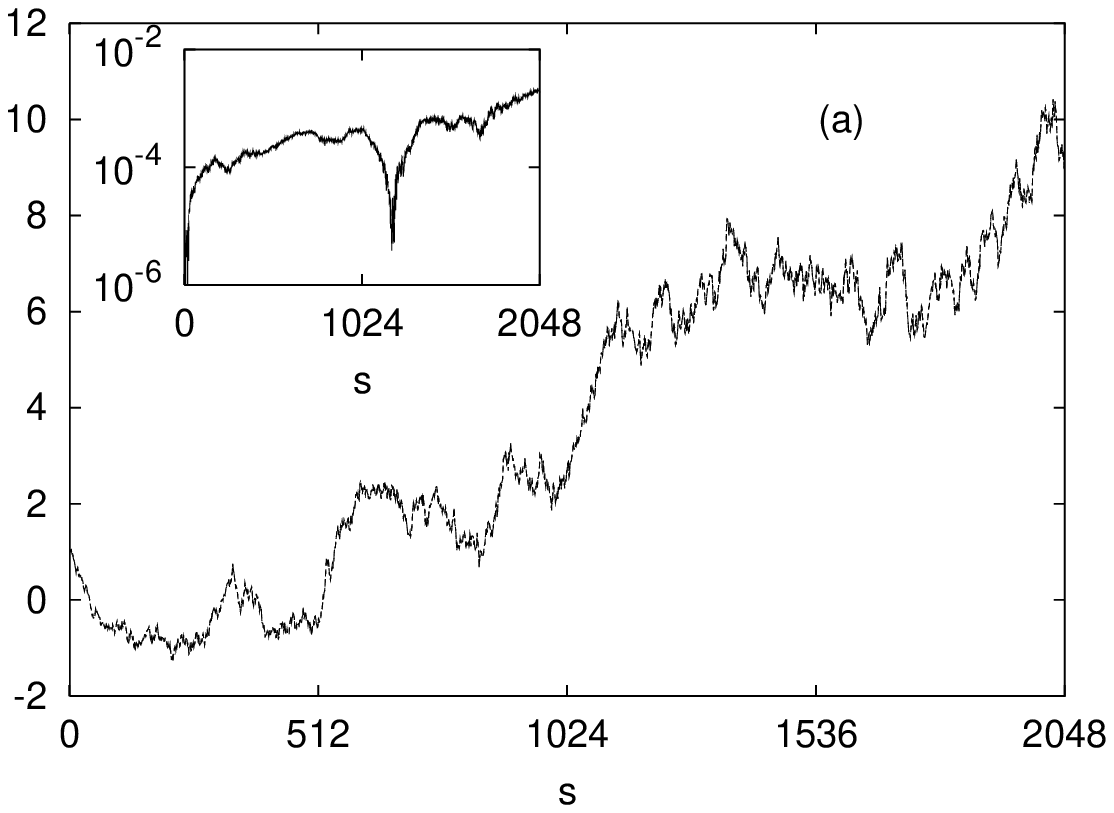}\hspace{-.3cm}
 \includegraphics[scale=.6,draft=false]{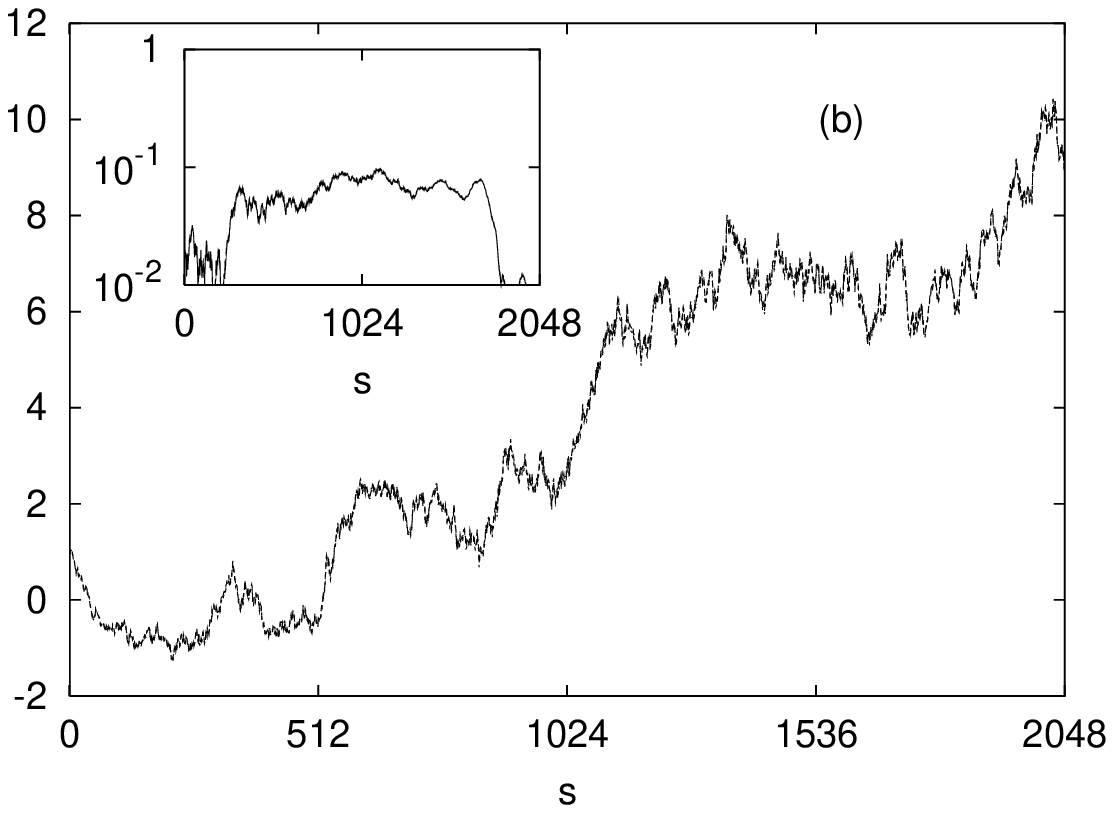}}
 \caption{
(a) $\int_0^s\d s'\, \int\!\!\d{\bm y}\, P({\bm y},s'|{\bm x},t)\phi({\bm
y},s')$ and $\int\!\!\d{\bm y}\, P({\bm y},s|{\bm x},t)\theta({\bm y},s)$ as a
function of time~$s$. Time is expressed in unit of time steps (longer
integration times display the same features). The difference is detectable only
looking at absolute errors, here shown in the inset.
(b) $\int\!\!\d{\bm y}\, P({\bm y},s|{\bm x},t)\theta({\bm y},s)$ obtained
integrating the propagator and by integrating $10^6$ particles initially
distributed according to~$P({\bm y},t|{\bm x},t)$. In the inset, the absolute
error.}
 \label{fig1}
\end{figure}
The quality of the integration can be tested using the following relation
\begin{equation}
 \int_0^s\!\!\d s' \int\!\!\d{\bm y}\;
 P({\bm y},s'|{\bm x},t)\,\phi({\bm y},s') =
 \int\!\!\d{\bm y}\; P({\bm y},s|{\bm x},t)\,\theta({\bm y},s)\;,
 \label{eq:8}
\end{equation}
which stems from~(\ref{eq:4.2}) and~(\ref{eq:5}).  In Fig.~\ref{fig1}a we show
both sides of~(\ref{eq:8}), the quality of the integration is rather good.

We have also performed Lagrangian simulations, i.e. we have integrated particle
trajectories evolving backward in time according to Eq.~(\ref{eq:1}).  For the
integration we used an Euler-It\^{o} scheme, and the particle velocity has been
obtained by means of a bilinear interpolation.  In Fig.~\ref{fig1}b we show the
r.h.s of~(\ref{eq:8}) evaluated with the propagator and with the Lagrangian
trajectories the final condition of which have been set according to the
propagator distribution~$P({\bm y},t|{\bm x},t)$.  We recall that in the limit
of infinite particles the propagator is exactly recovered. The good agreement
of Fig.~\ref{fig1}b reflects the fact that, although pseudo-spectral methods
are not suited to preserve the positivity of the propagator, the presence of
small negative regions is not severely penalizing. Indeed, a closer inspection
of the propagator shows that the negative values are limited to small amplitude
oscillations where~$P({\bm y},s|{\bm x},t)$ is vanishingly small.


\subsection{Frontogenesis in passive scalar advection}

\begin{figure}[tb]
 \centerline{
 \includegraphics[height=6.2cm,width=6.2cm,draft=false]{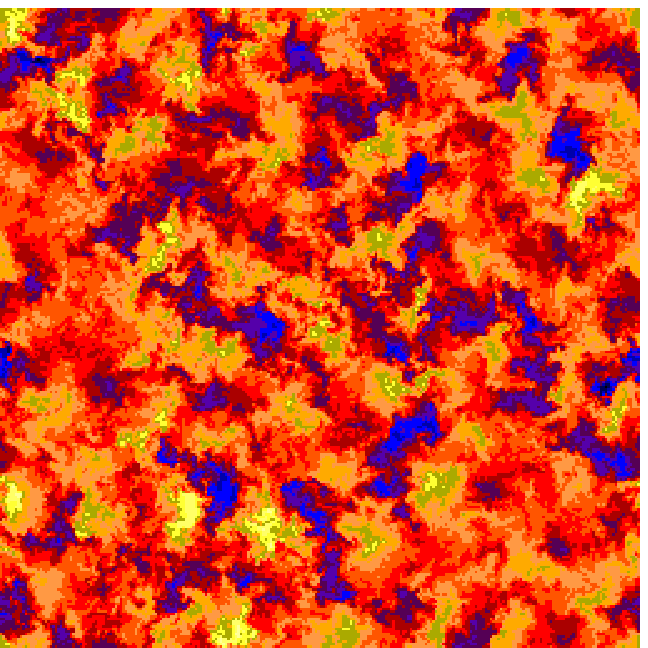}\hfill
 \includegraphics[height=6.2cm,width=6.2cm,draft=false]{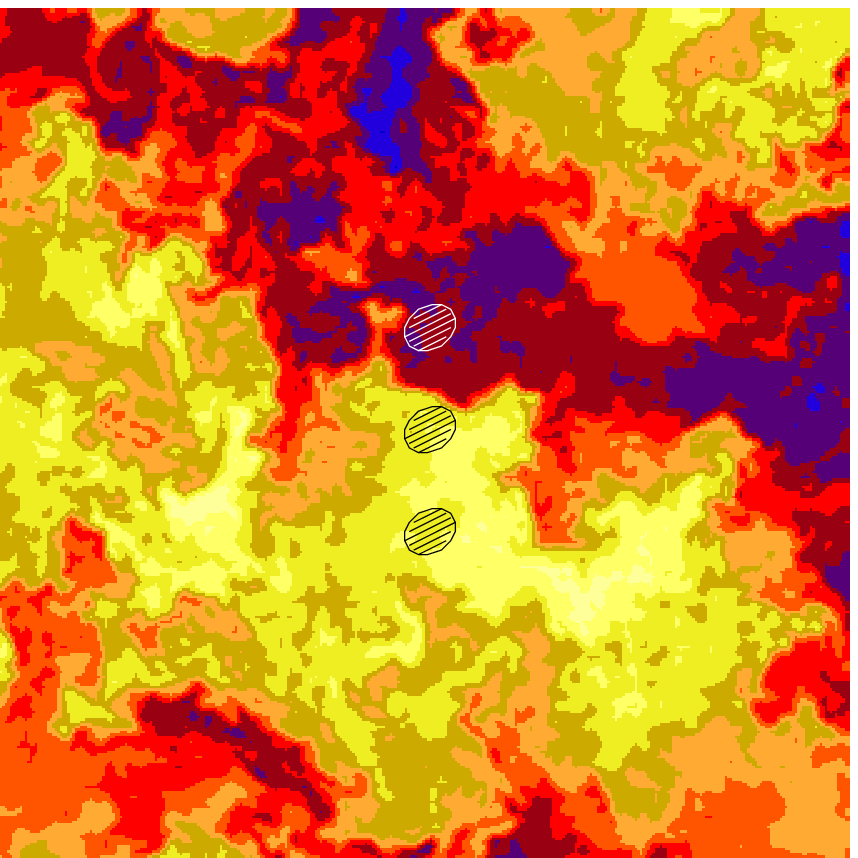}}
 \caption{
Left\,: Typical snapshots of the scalar field $\theta$.  Note the presence of
sharp fronts separating  large regions in which the scalar assumes close values.
Right\,: Close-up of a region containing a front. Across the upper and the
middle spot there is a front, whereas the middle and lower one lie in a
plateau. The distance between consecutive spots is larger than the diffusive
scale~$L_{\kappa}$, but smaller than the injection scale~$L_{\phi}$. In this
simulation $L_{\kappa} \approx 2$ , $L_\phi \approx 170$, the spot separation
and diameter are~$\approx 25$ and~$\approx 15$, respectively.  Lengths are
expressed in grid points.}
 \label{fig2}
\end{figure}

A striking and ubiquitous feature of passive scalar turbulence is the presence
of fronts (also called ``cliffs'' or ``sheets''), i.e.  regions where the
scalar has very strong variations separated by large regions (``ramps'' or
``plateaux'') where scalar fluctuations are weak (see
Fig.~\ref{fig2})~\cite{CLMV00,CLMV01,DSCV94,L88,KRS91,MW98,MWAT01,P94,CK98}.

\begin{figure}
 \centerline{\includegraphics[scale=1.2,draft=false]{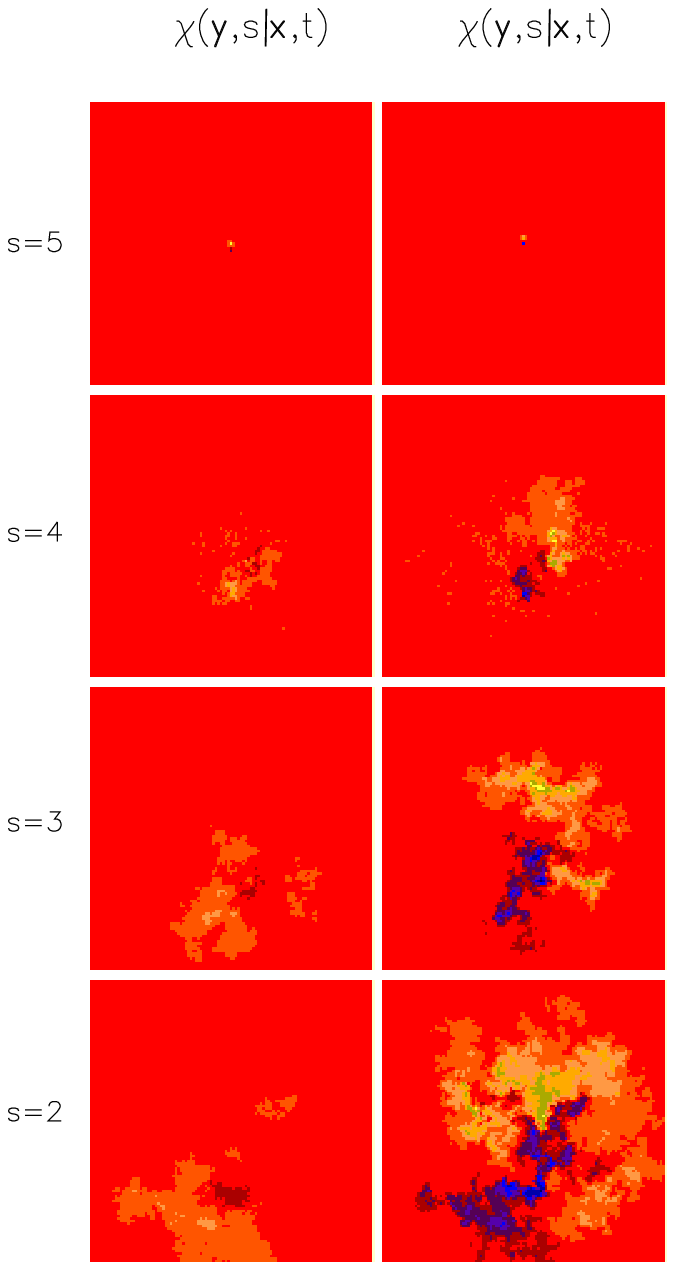}}
 \caption{
From top to bottom\,: backward evolution of~$\chi({\bm y},s|{\bm x},t)$ for
${\bm x}$ and ${\bm r}$ starting in a plateau (first column) and across a front
(second column), see the right panel of Fig.~\ref{fig2}. Colors are coded
according to the intensity of the field~$\chi$, yellow is for positive values
and blue for negative ones. At each time, the intensity is normalized according
to the maximum of the fields in absolute value.  The relatively smaller
intensity on the first column is due to the fast mixing leading to strong
cancellations between the positive and negative parts.  Time is in eddy
turnover times, the total number of time steps is~$2^{14}$.  To compare with
figure~\ref{fig2}, here the panel is $900 \times 900$~grid points.}
 \label{fig3}
\end{figure}

The genesis of fronts and plateaux is best understood in terms of
particle trajectories\,: to trace back the build-up of large and small
scalar difference we study the evolution of the propagator
\begin{equation}
 \chi({\bm y},s|{\bm x},{\bm x}+{\bm r},t) =
 P({\bm y},s|{\bm x}+{\bm r},t)-P({\bm y},s|{\bm x},t)\;,
\end{equation}
which is related to the scalar difference by the formula
\begin{equation}
 \theta({\bm x}+{\bm r},t)-\theta({\bm x},t) =
 \int_0^t\!\!\d s \int\!\!\d{\bm y}\;
 \chi({\bm y},s|{\bm x},{\bm x}+{\bm r},t)\,\phi({\bm y},s)\;.
\end{equation}

Notice that $\chi$~evolves backward according to Eq.~(\ref{eq:4.2}), with the
final condition $\chi({\bm y},s|{\bm x},{\bm x}+{\bm r},t) = \delta({\bm
y}-{\bm x}-{\bm r})-\delta({\bm y}-{\bm x})$.

The numerical procedure was as follows.  After the integration of
Eqs.~(\ref{eq:5}) and~(\ref{eq:6}) over five eddy turnover times (the typical
time-scale of large-scale motion) we choose ${\bm x}$ and~${\bm r}$ such
that~${\bm x}$, ${\bm x}+{\bm r}$ are on a front or a plateau, respectively
(see the right panel of Fig.~\ref{fig2}).  Then $\chi$ is integrated backward
in time.  In Fig.~\ref{fig3}, we show four snapshots of the backward evolution
of the field~$\chi$.  Already at a first glance the evolution of~$\chi$ appears
very different for the two final conditions\,: the blobs starting inside a
plateau (first column) experience a strong mixing, while blobs lying initially
across a front mix very poorly remaining compact and far aside.  This is the
basic mechanism for the formation of intense structures in passive scalar
turbulence (for a related theoretical study see~\cite{BL98}).


\section*{Acknowledgments}
\label{sec:5}

We are grateful to G.~Boffetta, S.~Musacchio, and M.~Vergassola for several
useful discussions and suggestions.  M.C. has been supported by the EU under
the contract HPRN-CT-2000-00162.  A.C. acknowledges the EU contract
HPRN-CT-2002-00300. Numerical simulations have been performed at IDRIS
(project~021226).


\end{document}